\documentclass[usenatbib]{mnras}
\usepackage{times}
\usepackage{color}
\usepackage{tensor}
\usepackage{multirow}
\usepackage{graphicx}
\usepackage{amsmath}
\usepackage{natbib}
\usepackage{enumerate}
\usepackage{bm}
\usepackage{hyperref}
\usepackage{amssymb}

\usepackage[english]{babel}
\usepackage{mathptmx}
\usepackage{tabularx}
\usepackage{placeins}
\usepackage{feynmp}
\usepackage{caption}
\usepackage{float}

\def\lsim{\mathrel{\rlap{\lower3pt\hbox{$\sim$}}
    \raise1pt\hbox{$<$}}}                
\def\gsim{\mathrel{\rlap{\lower3pt\hbox{$\sim$}}
    \raise1pt\hbox{$>$}}}                

\renewcommand{\vec}[1]{\bm{#1}}

\begin{document}
\pubyear{2017}
\title [Beamed emission and the ionising background]{Large-scale fluctuations in the cosmic ionising
  background:\\ the impact of beamed source emission }
\author[Suarez \& Pontzen]{Teresita Suarez,  Andrew Pontzen
  \\
  {Department of Physics and Astronomy, University College
    London,
    London WC1E 6BT} \\
 }

\date{ 26 July 2017}
\maketitle

\begin{abstract}
  When modelling the ionisation of gas in the intergalactic medium
  after reionisation, it is standard practice to assume a uniform
  radiation background. This assumption is not always appropriate;
  models with radiative transfer show that large-scale ionisation rate
  fluctuations can have an observable impact on statistics of the
  Lyman-alpha forest.  We extend such calculations to include beaming
  of sources, which has previously been neglected but which is
  expected to be important if quasars dominate the ionising photon
  budget. Beaming has two effects: first, the physical number density
  of ionising sources is enhanced relative to that directly observed;
  and second, the radiative transfer itself is altered. We calculate
  both effects in a hard-edged beaming model where each source has a
  random orientation, using an equilibrium Boltzmann hierarchy in
  terms of spherical harmonics.  By studying the statistical
  properties of the resulting ionisation rate and {\sc Hi} density
  fields at redshift $z\sim 2.3$, we find that the two effects
  partially cancel each other; combined, they constitute a maximum
  $5\%$ correction to the power spectrum $P_{\mathrm{HI}}(k)$ at
  $k=0.04 \, h/\mathrm{Mpc}$. On very large scales
  ($k<0.01\, h/\mathrm{Mpc}$) the source density renormalisation
  dominates; it can reduce, by an order of magnitude, the contribution
  of ionising shot-noise to the intergalactic {\sc Hi} power spectrum.
  The effects of beaming should be considered when interpreting future
  observational datasets.
\end{abstract}

\vspace{1.5cm}

\section{Introduction}

Intergalactic neutral hydrogen can be detected in the spectra of
background quasars; absorption at the rest-frame Lyman-$\alpha$
transition gives rise to a ``forest'' with hundreds of distinct
absorption lines corresponding to neutral hydrogen at different redshifts
\citep{Weymann:1981aa}. On small scales, between $1$ and $40$ $h^{-1}$
Mpc comoving, the forest can be used as a statistical tracer of the
distribution of matter \citep{Viel:2005aa}.

In fact hydrogen in the intergalactic medium (IGM) at $z<5$ is highly ionised
by ultraviolet (UV) background radiation produced by stars and quasars
\citep{Croft:2004aa,Viel:2005aa}, leaving only a trace of {\sc
  Hi}. This UV background is therefore an essential element in
simulations of the forest \citep{Cen:1994aa}.  When
modelling Lyman-$\alpha$ absorption, the neutral hydrogen density is
assumed to be in ionisation equilibrium with a uniform ionising
background
\citep[e.g.][]{Katz:1996aa,McDonald:2003aa,Haehnelt:2001aa,Croft:2004aa}.
Theoretical and observational arguments both show that this assumption
can fail in various limits and a fluctuating UV background ought at
least in principle to be included in analyses of the forest
\citep{Maselli:2005aa}.

Recently some attention has been devoted to understanding
{\sc Hi} fluctuations on scales approaching the mean-free-path of an
ionising photon
\citep{Pontzen:2014aa,Gontcho14,Pontzen14b,BOSSLya17}. In this
large-scale limit, the correlation of {\sc Hi} with cosmological
density progressively weakens and eventually reverses sign because the
clustering of the radiation field becomes stronger than the clustering
of intergalactic hydrogen. Additionally, if quasars contribute
significantly to the
photon production budget, an uncorrelated shot-noise component is added
to the power due to their intrinsic rarity. 

A number of factors have been neglected from radiative transfer
calculations to date, however. These include beaming of sources,
variable heating from high-frequency photons, and time dependence. In
this paper we tackle the first of these simplifications and explore
the effect of quasar beaming on the shot-noise contribution to the
large-scale diffuse {\sc Hi} power spectrum.  We estimate the
correction to the radiation fluctuations when emission is not
isotropic but beamed for a random distribution of quasars. 

The plan for the remainder of this paper is as follows. In Sec.~2
we derive the emissivity power spectrum accounting for a population of
sources with fixed beam widths but random orientations. In Sec. 3 we
discuss the radiation transfer equation appropriate for this
distribution (with further detail in Appendix A). We present
the resulting power spectrum of the radiation and {\sc Hi}
fluctuations in Sec. 4 and summarise in Sec. 5.


\section{Fluctuations in the emissivity}
\label{spherical_harmonics}

\subsection{Correcting the number density of sources $\bar n$} \label{sec:corr-numb-dens}
\label{section_geometry}

\begin{figure}
\centering
\includegraphics[width=0.4\textwidth]{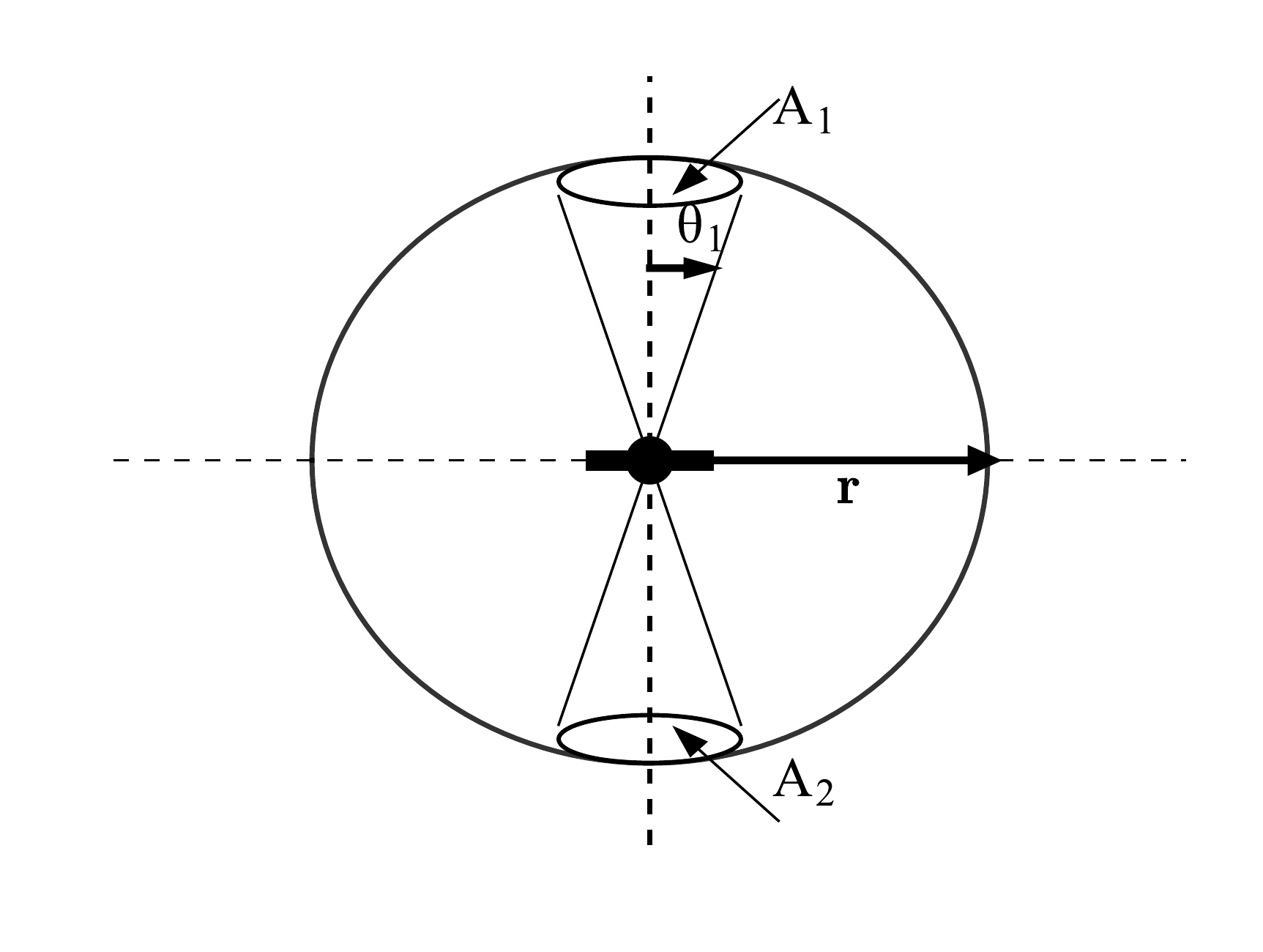}
\caption{Geometry of our source model, which represents a quasar with
  variable beam width. The probability of detecting a quasar is
  proportional to the beam area $A_1$ and $A_2$. These in turn are
  determined by the opening angle, here parameterised by $\theta_1$
  with $0<\theta_1\leq \pi/2$. The isotropic case is recovered when
  $\theta_1=\pi/2$.}
\label{quasar_geometry}
\end{figure}

The simplest effect of source beaming is that the underlying number
density $\bar n$ is no longer directly measured by observations.  
The observed number density $\bar n_{\mathrm{obs}}$ must be corrected
for the probability of being detected. This
probability is given by the area of the emission (the beam) divided by
the total area of a sphere, assuming a random orientation. We assume a
hard-edged beam with opening angle $2 \theta_1$; see Fig.
\ref{quasar_geometry}. The chance of any given quasar to be seen is
then
\begin{equation}
p(\mathrm{seen}) = \frac{(A_1 + A_2)}{4\pi r^2}  =1-\cos \theta_1\textrm{,}
\end{equation}
in which $A_1$ and $A_2$ are the areas of two axisymmetric beams.  The
isotropic case is recovered when the angle $\theta_1$ is equal to
$\pi/2$. In the limit that $\theta_1$ approaches zero, the
emission becomes a pencil-beam and the likelihood of observation
becomes extremely small.

The number of density sources we detect, $\bar n_{\mathrm{obs}}$, is 
the true mean density $\bar n$ times the probability for observing each one:
\begin{equation}
\bar n_{\mathrm{obs}}= \bar n(1-\cos\theta_1).
\label{seen_quasars}
\end{equation}
In
\citeauthor{Pontzen:2014aa} (2014; henceforth P14), it was assumed these two densities are
equal; for the results in this work, we fix $\bar n_{\mathrm{obs}}$
at the value estimated by P14, meaning that the
underlying density $\bar n$ varies. We emphasise that
$\bar n_{\mathrm{obs}}$ is itself highly uncertain, but that for the purposes
of understanding the effects of beaming it is simplest to keep it fixed.

\subsection{Definitions required for the emissivity derivation}
\label{geometry_model}

In the remainder of Section \ref{spherical_harmonics} we will calculate the effects of beaming
on the emissivity power spectrum from discrete sources.  A fraction of
photons comes from recombination of the IGM, but following the
approach of P14 we account for those through an appropriate additional
term in the radiative transfer equation (see Section
\ref{sec:radi-transf-meth}). Here, we can therefore focus on the
discrete sources alone.

We start from a rate of emission of photons in a narrow band at
frequency $\nu$, in a small volume around comoving position $\bm{x}$
and in an interval around the direction vector $\bm{n}$; this is
denoted $j_\nu(\vec{x}, \vec{n})$. As in P14, we simplify to a
frequency-averaged quantity $j(\vec{x},\vec{n})$ where
\begin{equation}
j(\vec{x}, \vec{n}) = \int j_\nu(\vec{x},\vec{n})\sigma_{\mathrm{HI}}(\nu)\mathrm{d}\nu\,  \textrm{.}
\label{eq:00j}
\end{equation}
The goal is to model fractional variations of $j$ around
its mean value $\langle j \rangle$, motivating the definition
\begin{equation}
\delta_j(\vec{x}, \vec{n}) = \frac{j(\vec{x},\vec{n})}{\langle j
  \rangle} - 1\textrm{.}
\end{equation}
We assume that variations on sufficiently large scales these can be related to
the cosmological matter overdensity $\delta_\rho$ multiplied by a
constant bias $b_j$, plus a Gaussian white-noise field
to represent shot-noise from the the rarity of sources.  The
variations of the emissivity, $\delta_{j}$, on large-scales is
therefore written:
\begin{equation}
\delta_j(\vec{x}, \vec{n}) = b_j\delta_\rho(\vec{x}) + \delta_{j,\mathrm{SN}}(\vec{x}, \vec{n}) \textrm{,}
\label{deltaj}
\end{equation}
where $\delta_{\rho}$ is the fractional matter overdensity at position
$\bm{x}$. According to Eq. \eqref{deltaj}, we need only consider
the component $\delta_{j,\mathrm{SN}}$ in the present work; by
construction all angle-dependence arises in the shot-noise term and the
radiation fluctuations that correlate with the cosmological density
field will not be altered by beaming. As a final simplification, P14
section II.C argues that the shot-noise contribution from galaxies is
negligible (owing to their very high number density) and we can assume
all contributions to $\delta_{j,\mathrm{SN}}$ arise from quasars.

In the remainder of this paper we will often need to work with
Fourier-transformed and spherical harmonic representations of
functions. For any function $F(\vec{x},\vec{n})$, these are defined respectively as
\begin{align}
\tilde{F}(\vec{k},\vec{n}) & \equiv \frac{1}{(2\pi)^{3/2}} \int \mathrm{d}^3
\vec{x}\, e^{-i\vec{k} \cdot \vec{x}}\, F(\vec{x},\vec{n})\,\textrm{
                             and}\label{eq:define-fourier-trans} \\
F^{\ell m}(\vec{x}) & \equiv \int \mathrm{d}^2 \vec{n}\,Y^{*}_{\ell
  m}(\vec{n}) F(\vec{x}, \vec{n})\,,\label{eq:define-sph-harmons}
\end{align}
where $Y^{*}_{\ell m}(\vec{n})$ is the complex conjugate spherical
harmonic basis function as defined in \cite{Varshalovich88}. The
spherical harmonic Fourier modes $\tilde{F}^{\ell m}(\vec{k})$ follow
by Fourier transforming Eq.~\eqref{eq:define-sph-harmons} or,
equivalently, taking spherical harmonics of Eq.
\eqref{eq:define-fourier-trans}.


\subsection{Emission of one quasar with a preferred alignment}
We want to understand the statistical properties of the $j{\bf(x, n)}$
field accounting for anisotropic emission from the
sources. To start, consider a single quasar of luminosity $L$ inside a
fixed volume $V$. Adopting at first an aligned coordinate system such
that $\theta$ gives the angle to the symmetry axis, and using the
geometry of Sec.~\ref{section_geometry}, we have
\begin{equation}
j_{\mathrm{aligned}}(\theta,\phi) =J
\begin{cases}
1\;\;\;\;\;\;0<\theta<\theta_1\\
0\;\;\;\;\;\theta_1<\theta<\pi-\theta_1\\
1\;\;\;\;\;\pi-\theta_1<\theta<\pi\\
\end{cases}
\textrm{,}
\label{eq:00dj}
\end{equation}
\noindent where $j_{\mathrm{aligned}}(\theta,
\phi)$ indicates the emissivity for the single quasar in our preferred
coordinate system, the constant $J$ is defined by $J = L/(4\pi
V(1-\cos\theta_1)) $, and
$\theta_1$ is the angle described in Section \ref{sec:corr-numb-dens}
and can have any value in the interval [$0,\pi/2$].

To proceed further, we decompose the function
$j_{\mathrm{aligned}}(\theta,\phi)$ into spherical harmonics
$j_{\mathrm{aligned}}^{\ell m}$ according
to Eq. \eqref{eq:define-sph-harmons}. 
For our aligned choice of coordinates, we only need to consider $m=0$
terms because of the cylindrical symmetry around the $\hat{z}$ axis.
In this case the spherical harmonic $Y_{\ell 0}(\theta, \phi)$ can be
written in terms of a Legendre polynomial,
$ Y_{\ell 0}(\theta,\phi) = \sqrt{(2l+1)/(4
  \pi)}P_{\ell}(\cos\theta)$. Using this result and the recursion
relations for $P_{\ell}$, we can express the emissivity variations as
\begin{equation}
J_{\ell} \equiv j_{\mathrm{aligned}}^{\ell 0} = J
\begin{cases}
 \sqrt{4\pi} (1-\cos \theta_1)  & \ell=0;\\
 \sqrt{\frac{\pi}{2\ell+1}}   \Big[1 - (-1)^{\ell+1}\Big] \Big[P_{\ell-1} -
 P_{\ell+1} \Big] & \ell>0 \textrm{,}
\end{cases}
\label{eq:anisotropic_emission}
\end{equation}

\noindent where for brevity we have written the Legendre
polynomials evaluated at $\cos \theta_1$, i.e. $P_{l-1} \equiv P_{l-1}(\cos
\theta_1)$  and $P_{l+1} \equiv P_{l+1}(\cos \theta_1)$.

\subsection{Emission of $N$ quasars with no preferred alignment}\label{sec:pj-power-spec}

So far we have derived the emissivity for a single quasar in a volume
$V$ in terms of the spherical harmonics coefficients. However, we now
need to look at the realistic case of $N$ quasars, each with a
beam pointing in an independent random direction. To achieve this we
first need to drop the assumption of a preferred coordinate system,
even for the single-quasar case $N=1$.

The spherical harmonic coefficients $j^{\ell m}_{N=1}$ for a single quasar
pointing in an arbitrary direction are related to the aligned spherical
harmonics via the Wigner $D$ matrix, which expresses a rotation by
Euler angles $\phi$, $\theta$, $\psi$:
\begin{equation}
\begin{split}
j^{\ell m}_{N=1} &= \sum_{m'} D^{\ell}_{mm'}(\phi, \theta, \psi)\;
j_{\mathrm{aligned}}^{\ell m'} \\
& =  D^\ell_{m0}(\phi, \theta, \psi)\; J_\ell  \textrm{.}\\
\end{split}
\end{equation}
We are thus only interested in the value of the $D$ matrix when $m'=0$, for
which case we have the identity
\citep{Varshalovich88}:
\begin{equation}
D^\ell_{m0}(\phi, \theta, \psi) = \sqrt{\frac{4 \pi}{2 \ell +1}} Y^*_{\ell
  m}(\theta, \phi)\textrm{.}
\end{equation}
The average emissivity over all possible beam alignment Euler angles is now
given by
\begin{align}
  \langle j_{N=1}^{\ell m} \rangle_{\phi,\theta,\psi} & \equiv \frac{1}{8\pi^2} \iiint \mathrm{d}\theta
                                              \mathrm{d}\phi \mathrm{d}\psi
                                              \sin\theta \,
                                                  D^\ell_{m0}(\theta,\phi,\psi)\,J_\ell
  \nonumber \\
                                            & =
\begin{cases}
J_0 & \ell=0\,;\\
0 & \ell \neq 0\,.
\end{cases}\label{eq:angle-averaged-j-one-quasar}
\end{align}

\noindent Next we need to calculate the two-point statistic
$\langle j_{N=1}^{\ell m} j_{N=1}^{*\ell' m'}\rangle$. Still using a
single quasar averaged over all possible directions we obtain
\begin{equation}
\begin{split}
\langle j_{N=1}^{\ell m} j_{N=1}^{*\ell' m'}\rangle_{\phi,\theta,\psi} = \frac{1}{2\ell+1}
\begin{cases}
J_\ell^2 &  \ell = \ell',\; m=m' \,; \\
0 & \textrm{otherwise,} 
\end{cases}
\end{split}\label{eq:J-power-N=1}
\end{equation}

\noindent where we used the orthogonality properties of the spherical
harmonics.  

If $N$ sources contribute, the angle-averaged emissivity
\eqref{eq:angle-averaged-j-one-quasar} is simply scaled up by
a factor $N$. However the generalisation of the two-point function requires a
more careful analysis. The total emissivity $j_{N}^{\ell m}$ in this case can
be decomposed as the sum of emissivity due to the individual sources,
\begin{equation}
j_N^{\ell m} = \sum_{i=1}^N j_{(i)}^{\ell m}\,,
\end{equation}
where $j_{(i)}^{\ell m}$ represents the emission due to the $i$th
source. When taking an average over all possible orientations, each
now has its own Euler angles $\phi_i, \theta_i, \psi_i$. Considering
the two-point function,
\begin{equation}
\langle j_N^{\ell m} j_N^{*\ell' m'} \rangle_{\{\phi_i,\theta_i,\psi_i\}} = \sum_{a=1}^N
\sum_{b=1}^N \langle j_{(a)}^{\ell m} j_{(b)}^{\ell'
  m'}\rangle_{\{\phi_i,\theta_i,\psi_i\}}\,,
\end{equation}
there are now two types of term. First, there are the single-source
terms where $a=b$. In these cases, the average over Euler angles
is no different from the $N=1$ case. Since there
are $N$ such terms with $a=b$, they contribute $N$ times the result in Eq.
\eqref{eq:J-power-N=1}. Second, there are  cross-quasar terms
where $a \ne b$. These terms involve separately integrating over the Euler angles
for both $a$ and $b$. The decoupled integrals are individually of the
form \eqref{eq:angle-averaged-j-one-quasar}; each cross-term (of which
there are $N^2-N$ in total) therefore
contributes $J_0^2$ when $\ell=0$ and zero otherwise. Putting together
the results above we find that
 \begin{equation}
 \langle  j_{N}^{\ell m}j_N^{*\ell'm'} \rangle_{\{\phi_i,\theta_i,\psi_i\}} = 
 \frac{1}{2\ell+1}
 \begin{cases}
 N^2J_0^2 & \ell=\ell'=m=m'=0\,;\\
 NJ_\ell^2 & \ell=\ell'\ne 0,\; m=m'\,;\\
0 & \text{otherwise.}
 \end{cases}
 \label{eq:j-power-fixed-N}
 \end{equation}
 
 \noindent
This is the final result for the case of a fixed number of $N$ quasars
inside a volume $V$.
 
\subsection{Putting it together: emissivity shot-noise power spectrum}

So far we have considered a case where the emissivity inside a fixed
volume $V$ with a known number of quasars $N$ is calculated. We now
need to introduce fluctuations in $N$. We expect to have
$\langle N \rangle = \bar n V$ quasars, where the average is over all
possible values of $N$ and $\bar n$ is the number density.  In the
Gaussian limit of Poisson statistics (which should be appropriate on
large scales), we also know that the variance in $N$ is given
by the relation
$\langle N^2 \rangle - \langle N \rangle ^2 = \bar{n}
V$. Consequently, the statistics for the emissivity $j_V$ in a fixed
volume but with varying $N$ are specified by
 \begin{equation}
 \langle  j_V^{\ell m}j_V^{*\ell'm'} \rangle = 
 \frac{1}{2\ell+1}
 \begin{cases}
 \left[\bar n V + (\bar n V)^2 \right] J_0^2 & \ell=\ell'=m=m'=0\,;\\
 \bar n V J_\ell^2 & \ell=\ell'\ne 0,\; m=m'\,;\\
0 & \text{otherwise,}
 \end{cases}
 \label{eq:jlmvol}
 \end{equation}
 where the average is now over all values of $N$, as well as over the
 Euler angles $\phi_i$, $\theta_i$, $\psi_i$ for each quasar. 

 To make the connection with the statistics of the
 $\delta_{j,\mathrm{SN}}$ field introduced in Eq. \eqref{deltaj},
 we now define the fractional variations in $j$ in a volume $V$ as
 $\delta_{j,V}$ where
\begin{equation}
\delta_{j,V}(\vec{n}) \equiv \frac{j_V(\vec{n})- \langle j_V \rangle}{\langle j_V \rangle } \textrm{,}
\end{equation}

\noindent which implies the spherical harmonic expansion is given by 

\begin{equation}
\delta^{\ell m}_{j,V} =  \frac{j_V^{\ell m} - \langle j_V^{\ell m} \rangle}{\langle j_V^{00}\rangle}\sqrt{4\pi} \textrm{.}
\end{equation}
Using this result alongside Eq. (\ref{eq:jlmvol}) we find that $\langle
\delta_{j,V}^{\ell m}\rangle = 0$ and 
\begin{equation}
\langle   \delta_{j,V}^{\ell m}\delta_{j,V}^{*\ell 'm'} \rangle = \frac{1}{\bar n V}\frac{4\pi}{2\ell+1}
\begin{cases}
(J_\ell/J_0)^2 & \ell=\ell',\;m=m'\textrm{;}\\
0 & \text{otherwise.}\
\end{cases}
\label{eq:deltavol}
\end{equation}

\noindent Finally, the expression for the statistics averaged over a
volume $V$ need to be related to the power spectrum of
$\delta_{j,\mathrm{SN}}$. We define the source shot-noise power
spectrum $P_{j,\mathrm{SN},\ell}(\bm{k})$ via
\begin{equation}
 \langle \tilde\delta_{j,\mathrm{SN}}^{\ell
   m}(\vec{k})\tilde\delta_{j,\mathrm{SN}}^{*\ell' m'}(\vec{k}') \rangle
 = P_{j,\mathrm{SN},\ell }(\bm{k}) \delta_\mathrm{D}(\bm{k} -
 \bm{k}')\delta_{\ell \ell'}\delta_{mm'}\textrm{,}
 \end{equation}
 \noindent where $\delta_\mathrm{D}$ is the Dirac delta
 function. To make contact between this required form and the
 derivation so far, one calculates the fluctuation averaged over a
 volume $V$:
\begin{align}
\delta_{j,V}^{\ell m} & \equiv \frac{1}{V} \int_V \mathrm{d}^3 x\,
\delta_{j,\mathrm{SN}}^{\ell,m}(\vec{x}) \nonumber \\
& = \frac{1}{(2 \pi)^{3/2} V} \int_V \mathrm{d}^3 x \int \mathrm{d}^3
  k\,e^{i \vec{k} \cdot \vec{x}} \tilde \delta_{j,\mathrm{SN}}^{\ell,m}(\vec{k})
\textrm{.}
\end{align}
Making the ansatz that $P_{j,\mathrm{SN},\ell}(k)$ is in fact
   independent of $k$ (as expected for shot-noise), we
 find that
\begin{equation}
\langle \delta_{j,V}^{\ell m} \delta_{j,V}^{*\ell'm'} \rangle =
\frac{1}{V} P_{j,\mathrm{SN},\ell} \delta_{\ell \ell'} \delta_{m m'}\textrm{.}
\end{equation}
By comparing with Eq. (\ref{eq:deltavol}) one obtains the final result:
\begin{equation}
P_{j,\textrm{SN},\ell}(k)= \frac{4\pi}{\bar n}\frac{(J_\ell/J_0)^2}{2\ell+1} \; \textrm{.}\label{eq:source-shotnoise-power}
\end{equation}
In the case of isotropic emission, $J_{\ell}=0$ for $\ell>0$ and the
result \eqref{eq:source-shotnoise-power} agrees with that from
P14.  Note that the shot-noise always
scales with the inverse of the mean density $\bar n$ (whether or not
the emission is isotropic).

\section{Radiative transfer method}\label{sec:radi-transf-meth}

In this Section we expand the P14 linearised radiative transfer
equation into a spherical harmonic Boltzmann hierarchy so that we can
use the directional source statistics derived in
Sec.~\ref{spherical_harmonics}.  Starting from the physical number
density of photons $ f(\vec{x}, \vec{n},\nu)$ at co-moving position
$\vec{x}$ traveling in direction $\vec{n}$ with frequency $\nu$,
P14 integrates over the frequency dependence $\nu$ by defining
\begin{equation}
f_{\mathrm{LL}}(\vec{x}, \vec{n}) = \int \mathrm{d} \nu \, f(\vec{x},
\vec{n}, \nu) \sigma_{\mathrm{HI}}(\nu),
\end{equation}
analogous to Eq.~\eqref{eq:00j}.  We approximate the radiation
and ionisation to be in equilibrium; this is a good approximation at
redshift $z\sim2.3$ \citep[e.g.][]{Busca:2013aa}. The fractional variations
around the mean value of $f_{\mathrm{LL}}$ are denoted
$\delta_{f_\mathrm{LL}}$; by linearising the Boltzmann equation, P14
obtained

\begin{equation}
\begin{split}
 \left[i(a\,\kappa_\mathrm{tot,0})^{-1}({\bf n\cdot k})+1\right] & \tilde \delta_{f_\mathrm{LL}}(\textbf{k, n}) = (1-\beta_{\mathrm{H\textsc {i}}}\beta_\mathrm{r})\tilde\delta_j (\textbf{k, n})\\
&+ \beta_{\mathrm{H\textsc{i}}}\beta_\mathrm{r}[\tilde\delta_{n_{\mathrm{H\textsc{i}}}}+\tilde\delta_\Gamma(\vec{k})]-\tilde\delta_{\kappa_\mathrm{tot}} \textrm{.}
\end{split}
\label{eq:perturbations}
\end{equation}
\noindent Here, $a$ is the cosmological scalefactor and
$\kappa_{\mathrm{tot}}$ an effective opacity to ionising photons
(comprised of both physical absorption and corrections from effects
such as redshifting). The mean effective opacity is given by
$\kappa_{\mathrm{tot},0} \equiv \langle \kappa_{\mathrm{tot}} \rangle$
while its fractional fluctuations are specified by
$\delta_{\kappa_{\mathrm{tot}}}$. The inverse of $a\,\kappa_{\mathrm{tot},0}$ gives
the effective comoving mean free path of an ionising photon, which can
be estimated to be 350\,Mpc at $z=2.3$ (see P14 Eq. 16).  The
dimensionless quantities $\beta_{\mathrm{HI}}$ and $\beta_r$ quantify
respectively the fraction of effective opacity resulting from physical
absorption in the IGM, and the fraction of {\sc Hi} recombinations
that result in an emission of a new ionising photon. The appearance of
$\beta_r$ is in fact accounting for ionising photons re-emitted
(isotropically) from the IGM itself as mentioned at the start of
Section \ref{geometry_model}. For full details see P14.

In the same way that Eq.~\eqref{deltaj} decomposes the
emissivity into shot-noise and cosmological terms, we can decompose the
radiation density fluctuations:
\begin{equation}
  \tilde\delta_{f_{\mathrm{LL}}}(\vec{k}, \vec{n}) = \tilde\delta_{f_{\mathrm{LL}},\mathrm{SN}}(\vec{k}, \vec{n}) + b_{f_{\mathrm{LL}}}(\hat{\vec{k}}\cdot\vec{n}) \tilde\delta_{\rho}(\vec{k})\,\textrm{,}\label{eq:fLL-decomposition}
\end{equation}
where $b_{f_{\mathrm{LL}}}$ is a scale- and direction-dependent bias.  At linear
order, $\tilde\delta_{f_{\mathrm{LL}},\mathrm{SN}}$ depends only on
$\tilde\delta_{j,\mathrm{SN}}$, not on the cosmological density
$\tilde\delta_{\rho}$. In the present analysis we revise only the
shot-noise component.

P14 assumes that $\delta_{j,\mathrm{SN}}$ is independent of
$\vec{n}$. In our case, we can no longer make this assumption. Instead
we write the direction vector
$\vec n = (\sin\theta\cos\phi, \sin\theta\sin\phi, \cos\theta)$ and,
to simplify the analysis, rotate the coordinate system\footnote{This
  freedom is available because we will ultimately consider only scalar
  quantities such as the ionisation rate and {\sc Hi}
  density. Statistical isotropy will then ensure the choice of
  $\hat{\vec{k}}$ direction in the analysis is irrelevant. Note that
  our special coordinate system in this section is independent of the
  alternative preferred system temporarily adopted in the early
  parts of Sec.~\ref{spherical_harmonics}.}  such that the
wavevector $\vec{k}$ lies along the $\hat{z}$ axis, i.e.
$\vec{k} = (0, 0, k)$. The product $\vec{n} \cdot \vec{k}$ appearing
in the radiative transfer equation then expands to
\begin{equation}
{\vec{n}\cdot \vec{k}} = k\,\cos\theta = \sqrt{\frac{4 \pi}{3}}\;k\; Y_{10}(\theta,\phi) 
\textrm{.}
\label{iden_sh}
\end{equation}
\noindent Using this result in Eq. \eqref{eq:perturbations} we
obtain (see Appendix \ref{App:AppendixA} for a detailed derivation):
\begin{equation}
\begin{split}
\tilde\delta_{f_\mathrm{LL},\mathrm{SN}}^{\ell m}  - \beta_{\mathrm{H\textsc{i}}} \delta^{00}_{f_\mathrm{LL},\mathrm{SN}}
\delta_{0\ell}\delta_{0m} + \frac{ik}{a\,\kappa_\mathrm{tot,0}} \times \hspace{3.4cm}
\\
\left\{ \tilde\delta_{f_\mathrm{LL},\mathrm{SN}}^{\ell-1,m}\sqrt{\frac{(\ell+m)(\ell-m)}{(2\ell-1)(2\ell+1)}} 
+  \tilde\delta_{f_\mathrm{LL},\mathrm{SN}}^{\ell+1,m}
\sqrt{\frac{(\ell+m+1)(\ell-m+1)}{(2\ell+1)(2\ell+3)}} \right\} \\
= ( 1 - \beta_{\mathrm{H\textsc{i}}}\beta_\mathrm{r})
\tilde\delta_{j ,\mathrm{SN}}^{\ell m}(\vec{k})\textrm{,}
\label{anisotropicBSH}
\end{split}
\end{equation}

\noindent where the explicit $\vec{k}$-dependence of
$\tilde{\delta}_{f_{\mathrm{LL}}}^{\ell m}$ has been omitted from the
expression for brevity. 
The expression can be rewritten schematically as
\begin{equation}
\sum_{\ell'} M_{m\ell \ell'}(\vec{k}) \delta_{f_{LL},\mathrm{SN}}^{\ell'
  m} (\vec{k}) = (1-\beta_{\mathrm{HI}} \beta_r) \delta_{j,\mathrm{SN}}^{\ell m}(\vec{k})\textrm{,}
\end{equation}
where 
$M_{m\ell \ell'}(\vec{k})$ contains the appropriate
equilibrium radiation transfer
coefficients from the left-hand-side of \eqref{anisotropicBSH}. We can invert the
linear relationship:
\begin{equation}
\tilde\delta_{f_{\mathrm{LL}},\mathrm{SN}}^{\ell m} (\vec{k}) = (1-\beta_{\mathrm{HI}}
\beta_{\mathrm{r}}) \sum_{\ell'} M^{-1}_{m\ell\ell'}(\vec{k}) \tilde\delta_{j,\mathrm{SN}}^{\ell'm}(\vec{k})\textrm{,}\label{eq:schematic-inversion}
\end{equation}
where the inverse matrix $M_m^{-1}$ satisfies
\begin{equation}
\sum_{\ell'}  M^{-1}_{m\ell\ell'} M^{\phantom{-1}}_{m\ell' \ell''} =
\delta_{\ell \ell''} \textrm{.}
\label{inverse_M}
\end{equation}
In our analysis, we are only interested in the overall radiation
intensity fluctuations --- i.e. the statistical properties of
$\tilde\delta_{f_\mathrm{LL}}^{00}$, or equivalently $\tilde\delta_{\Gamma}$. Therefore we
need to consider only the $m=0$ component of Eq.
\eqref{eq:schematic-inversion} since different $m$s do not couple to
each other.

\begin{figure}
\centering
\includegraphics[width=0.5\textwidth]{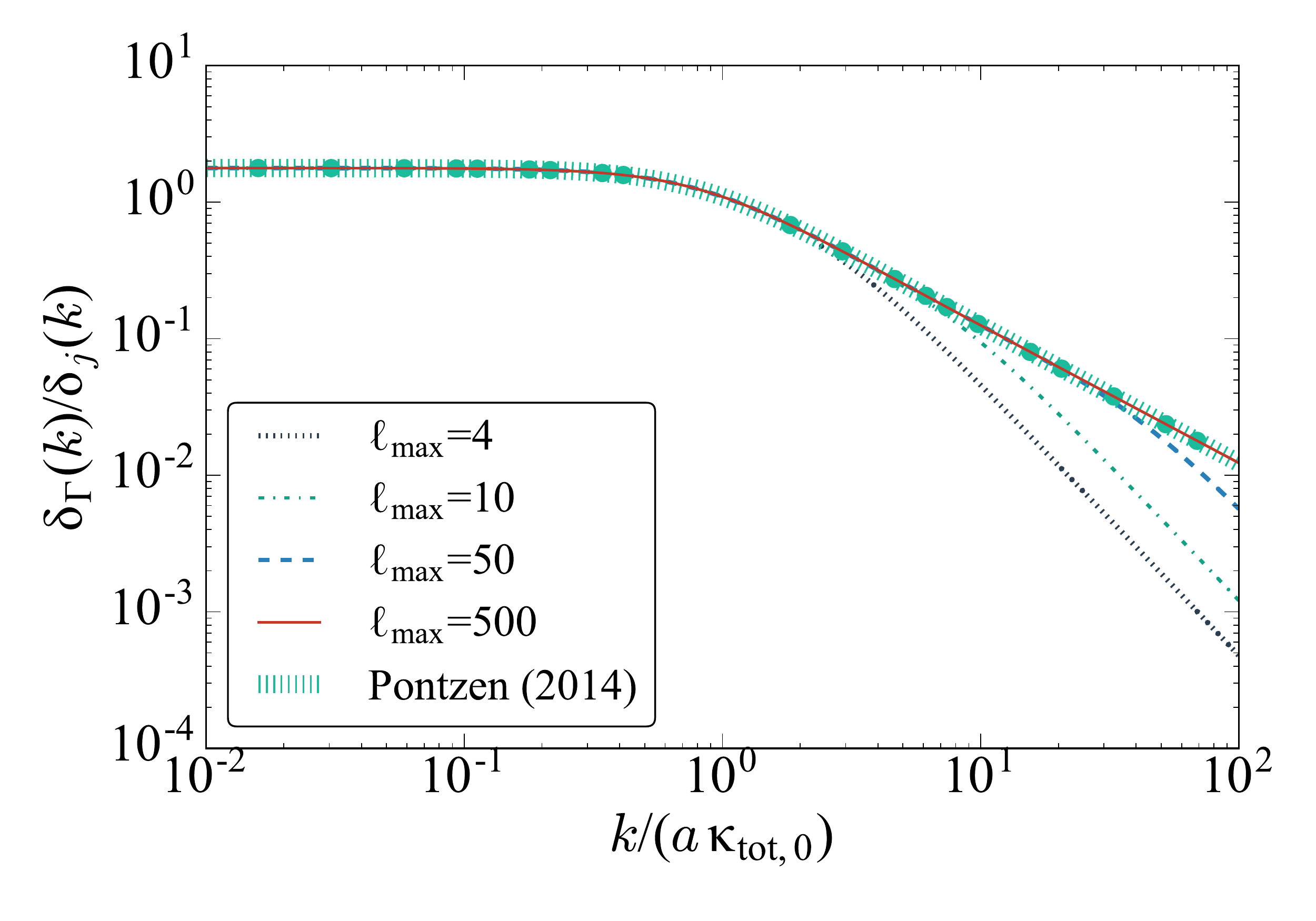}
\caption{ Demonstration of convergence of the numerical solution
  for shotnoise-induced ionisation rate fluctuations
  $\tilde\delta_{\Gamma,\mathrm{SN}}(\vec{k})$ as a fraction of the
  underlying source fluctuations $\tilde\delta_{j,\mathrm{SN}}(\vec{k})$. We
  choose an isotropic emission case where a closed-form analytic
  solution (shown by the shading) is known from
  \protect\cite{Pontzen:2014aa}. As $\ell_\mathrm{max}$ increases for
  test values from $4$ to $500$, our numerical hierarchy converges to
  this solution. }
\label{isotropic:eq00135}
\end{figure}

The next step is to solve the inversion \eqref{eq:schematic-inversion}
numerically. For these purposes the hierarchy must be truncated at
finite $\ell_{\mathrm{max}}$, so that we solve a
$\ell_{\mathrm{max}} \times \ell_{\mathrm{max}}$ matrix inversion for
each $k$. In practice choosing $\ell_{\mathrm{max}}$ requires a
convergence test to ensure that any results are insensitive to the
finite truncation. 

In the case of an isotropically radiating source, the solution for
$\tilde \delta_{\Gamma}$ was written in closed form by Eq. (30)
of P14, which can be rearranged (see Appendix \ref{App:AppendixA}) to
provide a test case that is illustrated in Fig.
\ref{isotropic:eq00135}. Specialising our Eq.
\eqref{eq:schematic-inversion} to the isotropic case corresponds to
setting $\tilde\delta^{\ell m}_j$ to zero for $\ell \ne 0$ and
$m \ne 0$.  The result for
$\delta_{\Gamma} = \delta_{f_{\mathrm{LL}}}^{00} / \sqrt{4 \pi}$ is
shown in Fig. \ref{isotropic:eq00135} as a function of
$k/(a\,\kappa_{\mathrm{tot,0}})$.

Consider first the closed-form solution from P14, shown by the shaded
band. The function shows how ionisation rate fluctuations trace
emissivity fluctuations on large scales (small $k$) -- the function
converges to a fixed, order unity value. On small scales (large $k$)
the fluctuations in the ionisation rate are suppressed: even a point
source of radiation can ionise extended regions of space, so the
small-scale ionisation rate fluctuations are damped. The
transition scale between these behaviours is set by the effective mean
free path $(a\,\kappa_{\mathrm{tot,0}})^{-1}$.

As $\ell_{\mathrm{max}}$ increases, our new hierarchy solution
correctly converges to the closed-form solution (dotted, dash-dotted,
dashed and solid lines respectively for $\ell_{\mathrm{max}} = 4$,
$10$, $50$ and $500$). The long wavelength limit (small $k$) is
completely insensitive to $\ell_{\mathrm{max}}$, while the highest $k$
modes are most sensitive. Note that while in this test case the
emission is isotropic, the actual radiation field is not; there is a
net flux of photons away from the plane wave peaks which defines a
preferred direction. This accounts for the sensitivity to
$\ell_{\mathrm{max}}$; as $k$ becomes large the spacing between peaks
becomes small compared to the mean free path of a photon. The true
photon distribution is sharply peaked in the $\hat{\vec{k}}$
direction and such sharp directionality requires a high
$\ell_{\mathrm{max}}$ for an adequate representation.

In the remainder of this paper we set $\ell_{\mathrm{max}}=500$; we
verified that increasing $\ell_{\mathrm{max}}$ to $1000$ did not
change our results.

\section{Results for ionisation rate and H{\sc i} power spectra}\label{sec:results}

We now have everything required to consider the power spectrum of radiation fluctuations for
different source beaming parameters. Again considering only the
shot-noise component, we can write 
\begin{equation}
P_{\Gamma,\mathrm{SN}}(k) = \frac{(1-\beta_{\mathrm{H\textsc{i}}} \beta_r)^2}{4 \pi} \sum_{\ell} (M^{-1}_{0 0 \ell}(k))^2 P_{j,\mathrm{SN},\ell}(k)\textrm{,}\label{eq:shotnoise-power-only}
\end{equation}
where $ P_{j,\mathrm{SN},\ell}(k)$ is given by Eq.
\eqref{eq:source-shotnoise-power} and is in fact independent of~$k$.
The term $M^{-1}_{0 0 \ell}(k)$ refers to the inverse matrix
  $M_{m\ell'\ell}^{-1}$ in Eq. (\ref{inverse_M}) with $m=0$ and
  $\ell'=0$; in general there is no closed analytic form so, as described in
  Section \ref{sec:radi-transf-meth},
  the matrix must be inverted numerically. 

\begin{figure}
\centering
\includegraphics[width=0.5\textwidth]{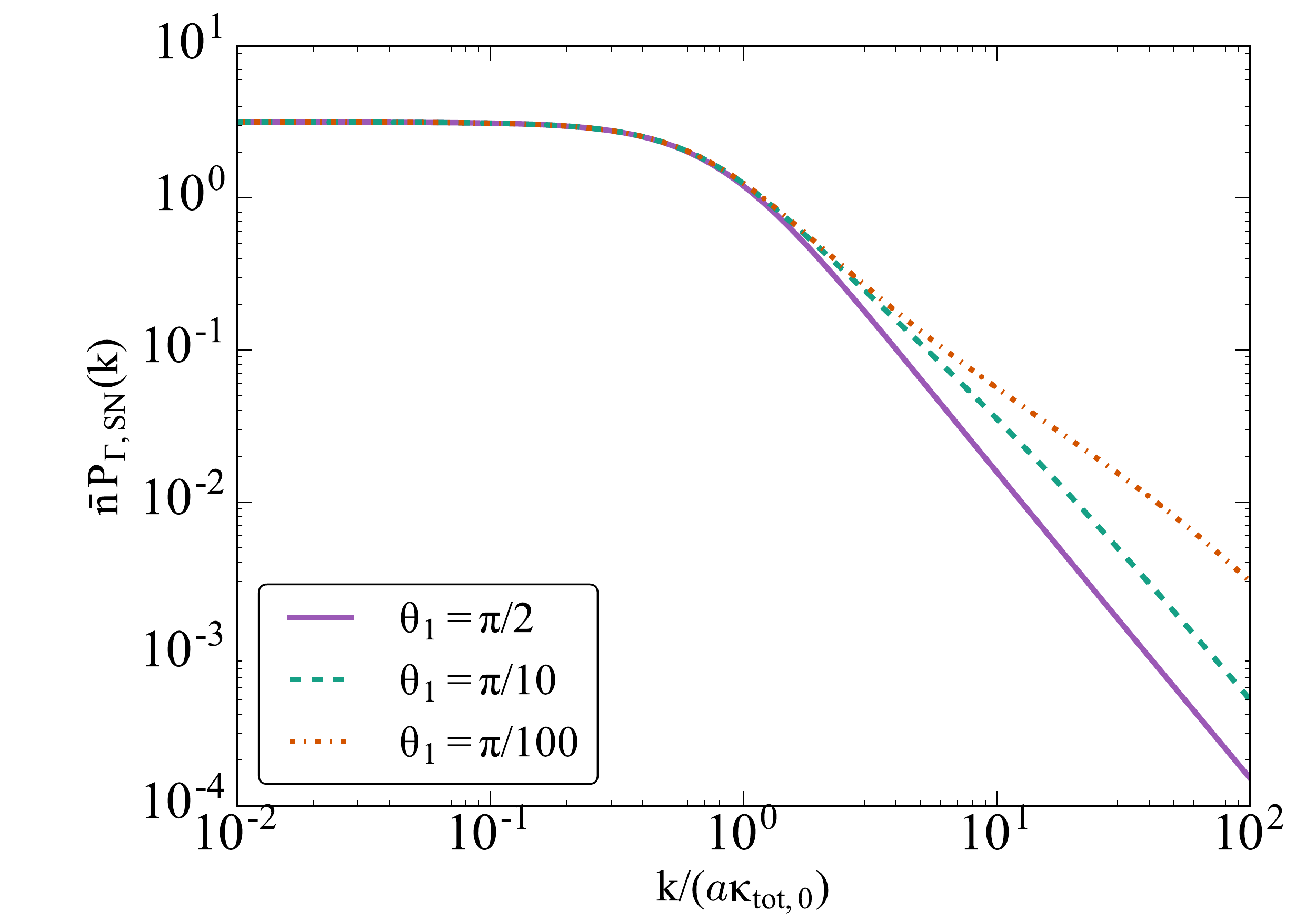}
\caption{ Shot noise power spectrum for radiation fluctuations
  $P_{\Gamma,\mathrm{SN}}(k)$, multiplied by the overall density of sources
  $\bar{n}$ (which removes the only dependence on $\bar{n}$). For the
  solid, dashed, and dash-dotted lines the opening angle for the
  radiation sources is $\pi/2$, $\pi/10$ and $\pi/100$
  respectively. The low $k$ radiation fluctuations are independent of
  the beam angle while at high $k$ the amplitude of fluctuations are
  increased for narrow beams. }
\label{power_spectrum_shot_noise}
\end{figure}

Figure \ref{power_spectrum_shot_noise} plots
$\bar{n}\,P_{\Gamma,SN}(k)$ against the wavenumber for different
values of the beam width $\theta_1 = \pi/2, \pi/10$ and
$\pi/100$. These correspond to opening angles of $180^\circ$,
$36^\circ$ and $3.6^\circ$ respectively; the last of these is
exceptionally narrow compared to observational estimates of $\gtrsim$ $30^{\circ}$
\citep[][]{TrainorSteidel13} and should be regarded as an extreme
upper limit on the magnitude of the correction.

Because $P_{j,\mathrm{SN},\ell}(k)$ scales inversely proportionally to the
source density $\bar{n}$, the product $\bar{n}\,P_{\Gamma,\mathrm{SN}}(k)$ is
independent of $\bar{n}$ and a function only of the beam
shape. Inspecting this product allows us to isolate and understand the
effect of the beaming.

The first case, $\theta_1=\pi/2$, is plotted with a solid line and
recovers the isotropic-emission solution (as previously illustrated by
Fig.~\ref{isotropic:eq00135}). As $\theta_1$ decreases (dashed and
dash-dotted lines respectively), the radiation emission is
increasingly tightly collimated. The effect of the beaming on
$\bar{n}\,P_{\Gamma,\mathrm{SN}}(k)$ is, however, confined to large
$k$. At low $k$, where the mean-free-path
$(a\kappa_{\mathrm{tot}})^{-1}$ is small compared to the wave under
consideration, beaming has no effect because the local ionising rate
scales proportionally to the total photon output of sources. For
smaller wavelengths (higher $k$), narrow beams lead to
higher-amplitude fluctuations in the ionising photon field.

Until now we have considered only the shot-noise contribution, but to
draw overall conclusions we need to put our revised radiation
shot-noise estimates back into the full calculation from
P14. Because the shot-noise is
uncorrelated with the cosmological fluctuations in the first-order
analysis, the power spectra add linearly:
\begin{equation}
\mathrm{P}_\mathrm{HI}(k) = b^2_\mathrm{HI}(k) \mathrm{P}_\rho(k) + \mathrm{P}_{\mathrm{HI},\mathrm{SN}}(k) \textrm{,}
\end{equation}
where $P_{\rho}(k)$ is the dark matter density power spectrum and
$b_\textsc{hi}$ is the linear relationship between H\textsc{i} density
and the total density as a function of scale, which is unchanged from
P14. Finally, because
$\delta_{n_{\mathrm{HI}},\mathrm{SN}} = - \delta_{\Gamma,\mathrm{SN}}$, the shot-noise
power spectra for {\sc Hi} and $\Gamma$ are equal and we have
\begin{equation}
\mathrm{P}_\mathrm{HI}(k) = b^2_\mathrm{HI} \mathrm{P}_\rho(\vec{k}) + \mathrm{P}_{\Gamma,\mathrm{SN}}(\vec{k}) \textrm{,}
\label{eq:final_total_power_spectrum}
\end{equation}
which allows us to use the result obtained in Eq.
\eqref{eq:shotnoise-power-only} for our beamed shot-noise estimates.

\begin{figure}
\centering
\includegraphics[width=0.5\textwidth]{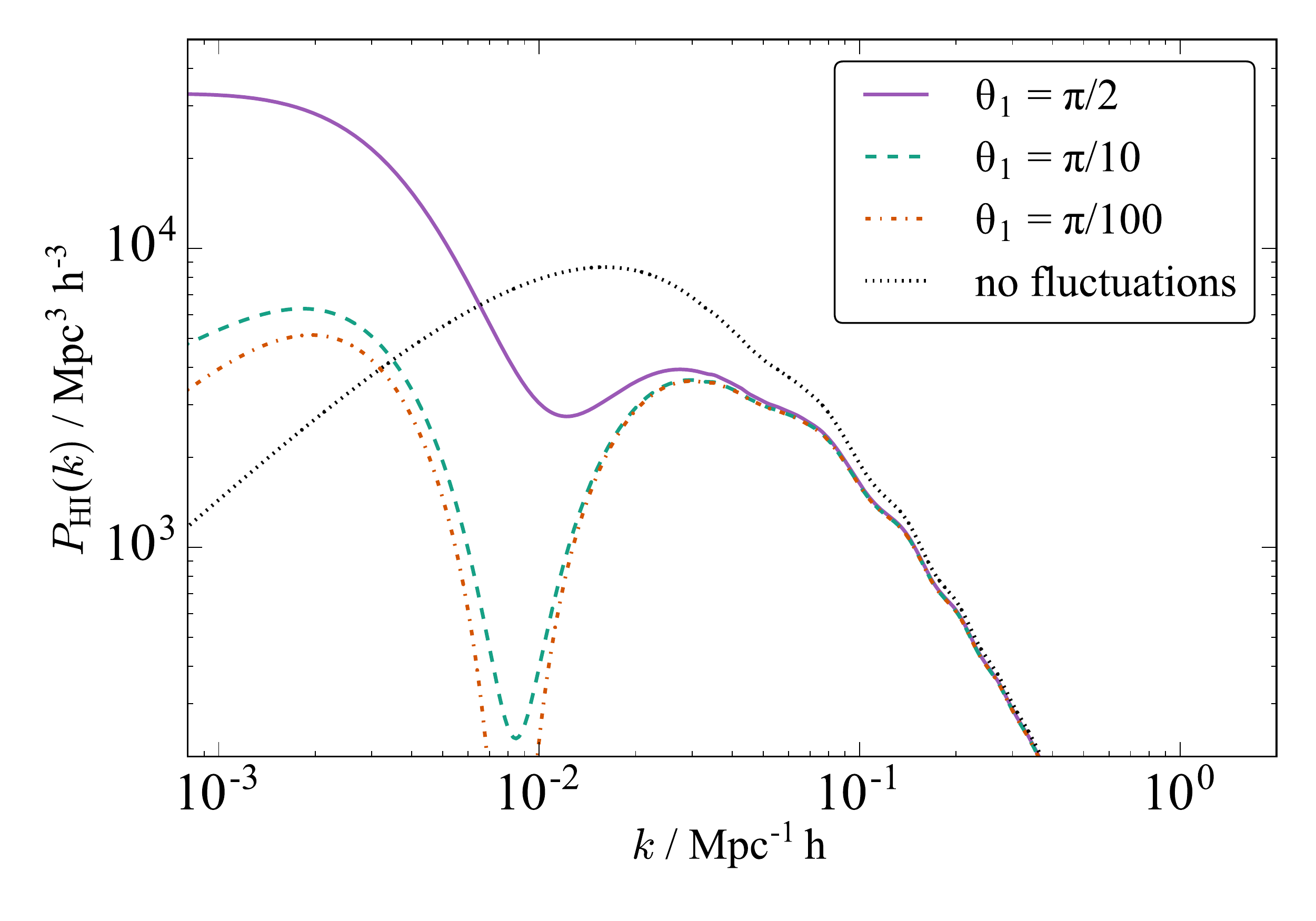}
\caption{ Power spectrum of the H\textsc{i} fluctuations defined by
  Eq. (\ref{eq:final_total_power_spectrum}), evaluated at
  redshift $z\simeq 2.3$ for a fixed observed source density of
  $\bar{n}_{\mathrm{obs}} = 10^{-4}\;h^3\;\mathrm{Mpc}^{-3}$, and for
  the same range of beam widths $\theta_1$ adopted in Fig.
  \ref{power_spectrum_shot_noise}. All other parameters are set to the
  defaults from \protect\cite{Pontzen:2014aa}. The dominant effect of
  changing $\theta_1$ arises from the rescaling of $\bar{n}$ to match
  $\bar{n}_{\mathrm{obs}}$; this can be seen on the largest scales
  (small $k$)
  where shot-noise effects dominate. The dotted line shows, for
  reference, the H{\sc i} power spectrum in the unphysical limit where
  there are no UV fluctuations. }
\label{total_power_spectrum}
\end{figure}

Figure \ref{total_power_spectrum} shows the total power spectrum for
the three values of $\theta_1$ previously adopted and a fixed density
of observed sources
$\bar{n}_{\mathrm{obs}}= 1\times10^{-4} \; h^3\;\mathrm{Mpc}^{-3}$, a
value adopted directly from P14. Note that, for our present
investigation, we assume that all sources have the same opening angle
(in particular ignoring the distinction between quasars and galaxies
in this respect).  Even if galaxies contribute comparable or larger
number of photons to the overall background, the shot-noise is still
strongly dominated by quasars (P14) and so this approximation is
likely valid.

As in P14, radiation can be approximated as near-uniform for the power
spectrum on scales below the mean-free-path (for
$k\gg 0.01\,h/\mathrm{Mpc}$); consequently the
corrections from the fluctuating ionisation always increase towards large
scales. On very large scales ($k<0.01\,h/\mathrm{Mpc}$), radiation
fluctuations actually dominate over H density fluctuations in the
\textsc{Hi} power spectrum. At the transition scale, there is a
characteristic dip where the radiation and H density
fluctuations approximately cancel. These basic features are
preserved when beaming is included.

As the source beams narrow, the results in
Fig. \ref{total_power_spectrum} show that the primary effect is to
reduce the amplitude of shot-noise fluctuations in the very large
scale regime ($k \ll 0.01\,h/\mathrm{Mpc}$) through the
renormalisation discussed in Sec. \ref{sec:corr-numb-dens}. Unlike the
direct effect of the beam, this observational correction to the
inferred number densities applies equally over all scales, making it
extremely significant in the low-$k$ regime where shot-noise
dominates.  However note that the constraining effect of current and
future surveys is quite poor on such extreme scales
\citep{Pontzen14b,BOSSLya17}. At scales $k \sim 0.04\, h/\mathrm{Mpc}$,
where reasonable observational precision can be expected in future
pipelines, the effects of beaming are considerably more modest
constituting a $\lsim 5\,\%$
correction. In this regime, the decreased power from
the source density renormalization is partially cancelled by the
increased power from the beaming itself (Figure
\ref{power_spectrum_shot_noise}). 

Because the effects are so strongly scale-dependent, and survey
sensitivities are also a steep function of scale, the observability of
beaming will be strongly dependent on details of observing strategy
and pipelines.

\section{Summary and discussion}
\label{conclusions}
We have constructed a model to treat fluctuations in the cosmological
UV background taking into account, for the first time, anisotropy due
to phenomena such as quasar beaming. To do that, we built upon the
monochromatic, equilibrium, large-scale description presented in
\cite{Pontzen:2014aa} but included angle-dependent emission terms.

We first introduced a correction for the observational bias that
individual quasars are less likely to be detected if they are tightly
beamed. This renormalises the underlying density of quasars in the
Universe. We then derived the emissivity shot-noise power spectrum
corresponding to a distribution of quasars each pointing in a
randomised direction. This required adopting an underlying beaming
model; for simplicity we used a hard-edged beam of fixed opening angle
$2\theta_1$ for the entire population (Fig.
\ref{quasar_geometry}). Finally, we rederived the radiative transfer
for the shot-noise taking into account the new angle-dependence. These
results are most naturally expressed in terms of a hierarchy of
spherical harmonic coefficients; because we are only
interested in the overall radiation intensity fluctuations we
ultimately took the $\ell=0$, $m=0$ component.

We solved the new hierarchy numerically and showed that, for
sufficiently large $\ell_\mathrm{max}$, the method converges to the
known isotropic-emission case when $\theta_1=\pi/2$ (see
Fig. \ref{isotropic:eq00135}) demonstrating the accuracy of the
method.  Then, we explored the effects of the quasar beam width
on the shot-noise (Fig. \ref{power_spectrum_shot_noise}).

The shot-noise power spectrum of radiation fluctuations
$P_{\Gamma, \textsc{sn}}$ in Fig. \ref{power_spectrum_shot_noise}
shows that the fluctuations are not sensitive to the beam
angle at low $k$. Fluctuation amplitudes do increase at high $k$ for
narrower beams; however the effect is modest even for extreme values
of the beam width $\theta_1$. 

When combining the new shot-noise solution with the cosmological
density fluctuations (Fig. \ref{total_power_spectrum}), we found
that the primary effect of beaming is in fact the first and simplest
one: the observational renormalisation of the underlying density of
bright sources. This affects all $k$-modes equally (scaling up and
down the overall contribution of shot-noise) and therefore is highly
significant on large scales (small $k$) where the shot-noise
potentially dominates over the cosmological signal.

This paper has focussed on clarifying one area where the effects of
radiative transfer on the Lyman-alpha forest were not known.  If future
pipelines lead to constraints on the magnitude of this effect
\citep[e.g.][]{Pontzen14b,BOSSLya17}, there are a number of other
possible influences that still require to be understood. For example,
time variability of sources and the effects of patchy heating
still need to be incorporated in a coherent framework and will be
tackled in future work.


\section*{Acknowledgments}
TS acknowledges support via studentships from CONACyT (Mexico) and
UCL.  AP is supported by the Royal Society. The authors would like to
thank Hiranya Peiris for many useful discussions and helpful comments
on a draft.

\bibliographystyle{mnras}
\bibliography{paper.bib}


\appendix
\section{Derivation of the radiative transfer hierarchy for
  anisotropic emission} \label{App:AppendixA} 

In this Appendix we present a derivation of our equilibrium 
hierarchy, Eq. (\ref{anisotropicBSH}), starting from the
Boltzmann equation \eqref{eq:perturbations}
which itself was previously derived in P14. As stated in the main text, we
choose a coordinate system in which the wavevector $\vec{k}$ lies
along the $\hat{z}$ axis, allowing us to rewrite $\vec{n} \cdot
\vec{k}$ in terms of $Y_1^0$; see Eq. \eqref{iden_sh}. 

To extract the spherical harmonic hierarchy, we expand  all angular
dependences; for any function $F$ one has
\begin{equation}
F(\vec{x}, \vec{n}) = \sum_{\ell', m'} F^{\ell' m'}(\vec{x}) Y_{\ell' m'}(\vec{n})\textrm{,}
\end{equation}
which is the inverse of the defining relation \eqref{eq:define-sph-harmons}.
We then multiply both sides of
Eq. (\ref{anisotropicBSH}) by $Y^*_{\ell' m'}(\vec{n})$ and
integrate over all angles $\vec{n}$. The left-hand-side becomes
\begin{equation}
\begin{split}
\mathrm{LHS} = \frac{ i k}{a
  \kappa_{\mathrm{tot},0}}\,\sqrt{\frac{4\pi}{3}}  &
\iint\,\mathrm{d}^2\vec{n}\, \sum_{\ell' ,m'} \tilde
\delta_{f_\mathrm{LL}}^{\ell' m'}({\bf k}) Y_{\ell' m'}(\vec{n})
Y_{1,0}(\vec{n}) Y_{\ell m}^*(\vec{n}) \\
 + & \delta_{f_\mathrm{LL}}^{\ell m}({\bf k})\textrm{,}
\end{split}
\label{l1:001}
\end{equation}
where to obtain the last term we have applied the orthogonality relation between spherical
harmonics.
The first term can be simplified by applying a
special case of the Wigner $3j$-symbol \citep{Varshalovich88}:
\begin{equation}
\begin{split}
\iint \mathrm{d}^2 \vec{n} \,Y_L^M(\vec{n})Y_1^0(\vec{n})Y_{L
  +1}^{*M}(\vec{n}) = \sqrt{\frac{3(L+M+1)(L-M+1)}{4\pi (2L+1)(2L+3)}}\textrm{.}
\end{split}
\label{wigner_identity}
\end{equation}
This identity can be used to calculate the integral for two values of
$\ell'$ in the sum given by \eqref{l1:001}, namely
$\ell' = \ell \pm 1$. The $\ell'=\ell+1$ case is obtained by a
relabelling of the indices whereas the $\ell'=\ell -1$ case is
obtained by taking the complex conjugate of Eq.
\eqref{wigner_identity}. By the triangle condition, integrals for any
other value of $\ell'$ vanish. Consequently we may write the LHS of
our expression as

\begin{equation}
\begin{split}
\mathrm{LHS} = &\frac{ik}{a\,\kappa_{\mathrm{tot},0}} \Biggr\{
\sqrt{\frac{(\ell +m)(\ell-m)}{(2 \ell-1)(2\ell+1)}} \tilde\delta_{f_{LL}}^{\ell-1,m}(\vec{k}) \quad
\\& \quad \quad \quad+  
\sqrt{\frac{(\ell+m+1)(\ell-m+1)}{(2\ell+1)(2\ell+3)}} \tilde\delta_{f_{LL}}^{\ell+1,m}(\vec{k}) \Biggr\} +
\tilde \delta_{f_{LL}}^{\ell m}({\bf k})\textrm{.}
\end{split}\label{eq:LHS-rad-transfer}
\end{equation}
We now turn to the right-hand-side of Eq.
(\ref{anisotropicBSH}), again multiplying by $Y^*_{\ell m}(\vec{n})$ and
integrating over all angles $\vec{n}$. Only $\tilde\delta_j$ has any
angular dependence on the RHS; the orthogonality of the spherical
harmonics picks out the coefficients $\tilde\delta^{\ell m}_j$ for
this term. For all other terms, only the $\ell=0$, $m=0$ case survives
the integration. The result is that
\begin{equation}
\begin{split}
\textrm{RHS} = & (1 - \beta_{\mathrm{HI}}\beta_r)  \tilde\delta_j^{\ell
  m} ({\bf k})  \\
& +\frac{\delta_{\ell 0} \delta_{m0}}{\sqrt{4
    \pi}}\left[\beta_{\mathrm{HI}}\beta_r\left(\tilde\delta_{n_{\mathrm{HI}}}(\vec{k})+\tilde\delta_\Gamma(\vec{k})\right)-\tilde\delta_{\kappa_{\rm
    tot}}(\vec{k})\right]\textrm{.}\label{eq:RHS-rad-transfer-intermediate}
\end{split}
\end{equation}
As expressed by Eq. \eqref{eq:fLL-decomposition}, we wish to
separate the radiation fluctuations that are correlated with the
cosmological density field $\delta_{\rho}$ from those that are caused
by shot-noise. To do so, we need to transform some of the terms on the
RHS which mix the two types of fluctuation as follows.

The terms $\tilde\delta_{n_{\mathrm{HI}}}$ and
$\tilde\delta_\Gamma$  do not have an angular dependence and their
relationships are therefore unchanged compared to P14:
\begin{equation}
\tilde\delta_{n_\mathrm{HI}} = \tilde\delta_{n_\mathrm{HI},u}-\tilde\delta_\Gamma ; \;\;\;\;\tilde \delta_{\kappa_\mathrm{tot}} = \beta_\mathrm{HI}\tilde\delta_{n_\mathrm{HI}} + \beta_\mathrm{clump}\tilde\delta_{\kappa_\mathrm{clump}}\
\textrm{.}
\label{identitiesf_LL}
\end{equation}
\noindent
The first of these relations arises from the fact that $n_\mathrm{H\textsc{i}}$ is
inversely proportional to the ionisation rate per H\textsc{i} atom,
recovering the completely uniform ionising background in the absence
of any radiative fluctuations, $\tilde\delta_\mathrm{n_{HI},u}$. The
effective opacity fluctuations $\delta_{\kappa_{\mathrm{tot}}}$
definition is a linear combination of the intergalactic medium
absorption fluctuations and the self-shielded clump opacity
fluctuations. 

We can use these relations to rewrite Eq.
\eqref{eq:RHS-rad-transfer-intermediate} in terms of
$\tilde \delta_{\Gamma}$, $\tilde \delta_{n_{\mathrm{HI}},u}$ and
$\tilde \delta_{\kappa_{\mathrm{clump}}}$. The ionisation rate fluctuations
$\tilde\delta_{\Gamma}$ are defined by the fluctuations in the photon
density $\tilde\delta_{f_{\mathrm{LL}}}$ via the relation 
\begin{equation}
 \delta_\Gamma = \frac{1}{4\pi} \int \mathrm{d}^2 n\,
  \tilde\delta_{f_{LL}}(\vec{k}, \vec{n}) = \frac{1}{\sqrt{4 \pi}} \tilde\delta^{00}_{f_{\mathrm{LL}}}(\vec{k}) \textrm{.}
\end{equation}
Consequently $\tilde\delta_{\Gamma}$ can be split into a correlated
and shot-noise component, with $\tilde \delta_{\Gamma,\mathrm{SN}} \equiv
\delta_{f_{\mathrm{LL}},\mathrm{SN}}^{00}/\sqrt{4\pi}$.

Using the above transformations, we find the RHS can be written
\begin{align}
\textrm{RHS} =\, & (1 - \beta_{\mathrm{HI}} \beta_r) \tilde \delta_j^{\ell
  m}(\vec{k}) + \beta_{\mathrm{HI}} \delta_{\ell0} \delta_{m0} \tilde
  \delta^{00}_{f_{\mathrm{LL}}}(\vec{k}) \nonumber \\
& +  \left( \beta_{\mathrm{HI}} \left(\beta_r-1\right) \tilde
  \delta_{n_{\mathrm{HI}},u}(\vec{k}) - \beta_{\mathrm{clump}} \tilde
  \delta_{\kappa_{\mathrm{clump}}} \right)\delta_{\ell 0}\delta_{m 0}\textrm{.}\label{eq:RHS-rad-transfer-separated}
\end{align}
We now apply the decomposition into cosmological and shot-noise
components for $\tilde \delta_{j}$ and
$\tilde \delta_{f_{\mathrm{LL}}}$ keeping only the shot-noise
contributions. (By linearity, the two types of contribution can be
treated independently and the cosmological terms are unchanged from
P14.) In particular, the terms $\delta_{n_{\mathrm{HI}},u}$ and
$\delta_{\kappa_{\mathrm{clump}}}$ are independent of radiation
fluctuations; consequently they have no shot-noise component and drop
out entirely. Combining Eqs. \eqref{eq:LHS-rad-transfer} and
\eqref{eq:RHS-rad-transfer-separated} one reaches the final
equation:
\begin{equation}
\begin{split}
\tilde\delta_{f_\mathrm{LL},\mathrm{SN}}^{\ell m}(\vec{k}) & - \beta_{\mathrm{H\textsc{i}}} \delta^{00}_{f_\mathrm{LL},\mathrm{SN}}(\vec{k})
\delta_{0\ell}\delta_{0m} 
\\
& \hspace{-0.5cm} + \frac{ik}{a\,\kappa_\mathrm{tot,0}} \times \left\{ \sqrt{\frac{(\ell+m)(\ell-m)}{(2\ell-1)(2\ell+1)}} \tilde\delta_{f_\mathrm{LL},\mathrm{SN}}^{\ell-1,m}(\vec{k})\right.
\\ 
& \quad+  \left. \sqrt{\frac{(\ell+m+1)(\ell-m+1)}{(2\ell+1)(2\ell+3)}}\tilde\delta_{f_\mathrm{LL},\mathrm{SN}}^{\ell+1,m}(\vec{k}) \right\} \\
& \hspace{3cm}= ( 1 - \beta_{\mathrm{H\textsc{i}}}\beta_\mathrm{r})
\tilde\delta_{j ,\mathrm{SN}}^{\ell m}(\vec{k})\textrm{,}
\end{split}
\end{equation}
which agrees with the expression \eqref{anisotropicBSH} provided in
the main text.

In Sec. \ref{sec:radi-transf-meth} we compared the solution to this
hierarchy with the known isotropic limit. For these purposes we need
to extract the shot-noise-only isotropic solution in a way that was not
 written in P14, although it is implicit in P14 Eq.
(38). Starting from P14's Eq. (30), we apply our Eq.
\eqref{identitiesf_LL} and again retain only those terms which are not
correlated with $\tilde\delta_{\rho}$. This yields
\begin{align}
\tilde\delta_{\Gamma,\mathrm{SN},\mathrm{iso}}(\vec{k}) &=
\frac{(1-\beta_{\mathrm{HI}} \beta_{\mathrm{r}})
  S(k)}{1-\beta_{\mathrm{HI}} S(k)} \tilde
\delta_{j\mathrm{,SN,iso}}(\vec{k})\textrm{,} \nonumber \\
\textrm{where } S(k) &= \frac{a \kappa_{\mathrm{tot},0}}{k} \arctan \frac{k}{a \kappa_{\mathrm{tot},0}}\textrm{,}
\end{align}
for the isotropic comparison case which is plotted as a shaded band in
Fig. \ref{isotropic:eq00135}.

\end{document}